\documentclass[
12pt,
prl,
a4paper,
amsmath,
superscriptaddress,
showkeys,
floatfix,
notitlepage,
]{revtex4-2}

\usepackage[font={stretch=1.2}, justification=Justified, format=plain]{caption}
\usepackage{graphicx}
\usepackage{dcolumn}
\usepackage{bm}
\usepackage{natbib}
\usepackage{xcolor}
\usepackage{siunitx}
\usepackage{upgreek}
\usepackage{newfloat}
\usepackage{xr}

\DeclareFloatingEnvironment[name={\bf Supplementary Fig.}]{suppfigure}

\makeatletter

\begin{document}

\preprint{APS/123-QED}

\title{Microscopic Geared Mechanisms}

\author{Gan Wang}
\affiliation{Department of Physics, University of Gothenburg, SE-41296, Gothenburg, Sweden}

\author{Marcel Rey}
\affiliation{Department of Physics, University of Gothenburg, SE-41296, Gothenburg, Sweden}
\affiliation{Institute of Physical Chemistry, University of Münster, 48149 Münster, Germany}

\author{Antonio Ciarlo}
\affiliation{Department of Physics, University of Gothenburg, SE-41296, Gothenburg, Sweden}

\author{Mahdi Shanei}
\affiliation{Department of Physics, Chalmers University of Technology, SE-41296, Gothenburg, Sweden}

\author{Kunli Xiong}
\affiliation{Department of Physics, University of Gothenburg, SE-41296, Gothenburg, Sweden}
\affiliation{Department of Materials Science and Engineering, Solid State Physics division, Uppsala University, SE-75121, Uppsala, Sweden}

\author{Giuseppe Pesce}
\affiliation{Department of Physics, University of Gothenburg, SE-41296, Gothenburg, Sweden}
\affiliation{Department of Physics E. Pancini, University of Naples Federico II, Complesso Universitario Monte Sant'Angelo, Via Cintia, 80126 Naples, Italy}

\author{Mikael K\"all}
\affiliation{Department of Physics, Chalmers University of Technology, SE-41296, Gothenburg, Sweden}

\author{Giovanni Volpe}
\affiliation{Department of Physics, University of Gothenburg, SE-41296, Gothenburg, Sweden}

\date{\today}

\begin{abstract}
The miniaturization of mechanical machines is critical for advancing nanotechnology and reducing device footprints. 
Traditional efforts to downsize gears and micromotors have faced limitations at around 0.1 mm for over thirty years due to the complexities of constructing drives and coupling systems at such scales. 
Here, we present an alternative approach utilizing optical metasurfaces to locally drive microscopic machines, which can then be fabricated using standard lithography techniques and seamlessly integrated on the chip, achieving sizes down to tens of micrometers with movements precise to the sub-micrometer scale. 
As a proof of principle, we demonstrate the construction of microscopic gear trains powered by a single driving gear with a metasurface activated by a plane light wave. 
Additionally, we develop a versatile pinion and rack micromachine capable of transducing rotational motion, performing periodic motion, and controlling microscopic mirrors for light deflection. 
Our on-chip fabrication process allows for straightforward parallelization and integration. Using light as a widely available and easily controllable energy source, these miniaturized metamachines offer precise control and movement, unlocking new possibilities for micro- and nanoscale systems.
\end{abstract}

\maketitle

\section{Introduction}\label{MainSection}

Geared mechanisms --- systems where interconnected gears transfer work and perform mechanical tasks --- have long mirrored the advancement of human technology. 
Their evolution spans from the wind and water mills of ancient times to the steam engines of the Industrial Revolution and on to modern automotive, aerospace, and robotics applications~\cite{radzevich1994handbook}. 
Current advances focus on miniaturizing these gears to micrometer scales~\cite{fernandez2020recent}. 
While enhancing material efficiency and reducing waste, this miniaturization also opens new possibilities for mechanizing and exploring a length scale that has largely remained elusive. 
For example, down-sizing geared mechanisms provides tools to gain a deeper understanding of microscopic phenomena such as friction~\cite{kim2007nanotribology, williams2006tribology} and surface interactions~\cite{kendall1994adhesion, maboudian1998surface}, while driving technological innovations such as high-performance microfluidic devices~\cite{ladavac2004microoptomechanical, neale2005all,maruo2006optically, maruo2007optically, baigl2012photo, butaite2019indirect, zhang2021reconfigurable} and  reconfigurable optical technologies~\cite{solgaard2014optical, gauthier2001optical, wu1997micromachining}.
Moreover, innovations in both manufacturing and powering these geared systems are also impacting fields such as microrobotics~\cite{hong2022magnetically, miskin2020electronically}, optical systems~\cite{babar2023self}, and force sensors~\cite{xu20243d}.

Efforts to miniaturize gears have primarily focused on creating individual micromotors --- microscopic objects capable of rotation. 
Various mechanisms have been explored to power these devices, including static~\cite{fan1989ic, tai1989ic,fennimore2003rotational,ghalichechian2008design} and AC electric fields~\cite{kim2014ultrahigh,shields2018supercolloidal,hong2024optoelectronically, shi2024dna}, magnetic fields~\cite{xia2010ferrofluids,zandrini2017magnetically,wang2017dynamic,wang2021light,liu2022creating}, and light fields~\cite{friese2001optically,kelemen2006integrated, bianchi2018optical,aubret2018targeted}. 
However, incorporating these micromotors into functional microscopic geared mechanisms has remained a significant challenge. Traditional semiconductor manufacturing methods for electrostatically driven gears~\cite{garcia1995surface, sniegowski1996surface} are hindered by the need for electric connectors, which occupy considerable space around each micromotor, limiting both miniaturization and parallelization. 
While far-field approaches such as AC electric, magnetic, and light fields allow further miniaturization, they present their own limitations. 
Methods employing AC electric~\cite{kim2014ultrahigh,shields2018supercolloidal,hong2024optoelectronically, shi2024dna} or magenetic fields~\cite{xia2010ferrofluids,zandrini2017magnetically,wang2017dynamic,wang2021light,liu2022creating} require motors to be made from specific materials, complicating their integration within commercial semiconductor microfabrication techniques and making it difficult to combine multiple mechanical components; moreover, these approaches often lack the ability to address gears individually~\cite{sitti2020pros}. 
Light-based methods~\cite{galajda2002complex, maruo2006optically, maruo2007optically}, like optical tweezers, are less constrained by material properties but require focused light beams, limiting their large-scale manipulation potential. 
While methods based on photo-generated electric fields~\cite{zhang2021reconfigurable} and light-driven chemical reactions~\cite{wang2024femtosecond, aubret2021metamachines} offer promising solutions, they typically lack flexibility and are restricted to specific chemical environments. 
As a result, microscopic geared mechanisms that overcomes all these limitations remains elusive.

Here, we address these challenges by fabricating geared mechanism driven by optical metasurfaces that operate under uniform illumination. Using silicon as the primary material ensures compatibility with standard photolithography, facilitating large-scale manufacturing. This approach creates a versatile platform for precise control and movement of geared functional devices, enabling unprecedented capabilities in micro- and nanoscale mechanical systems.

\subsection{Micromotors powered by optical metasurface}

\begin{figure}[h]
    \begin{center}
    \includegraphics[width=\textwidth]{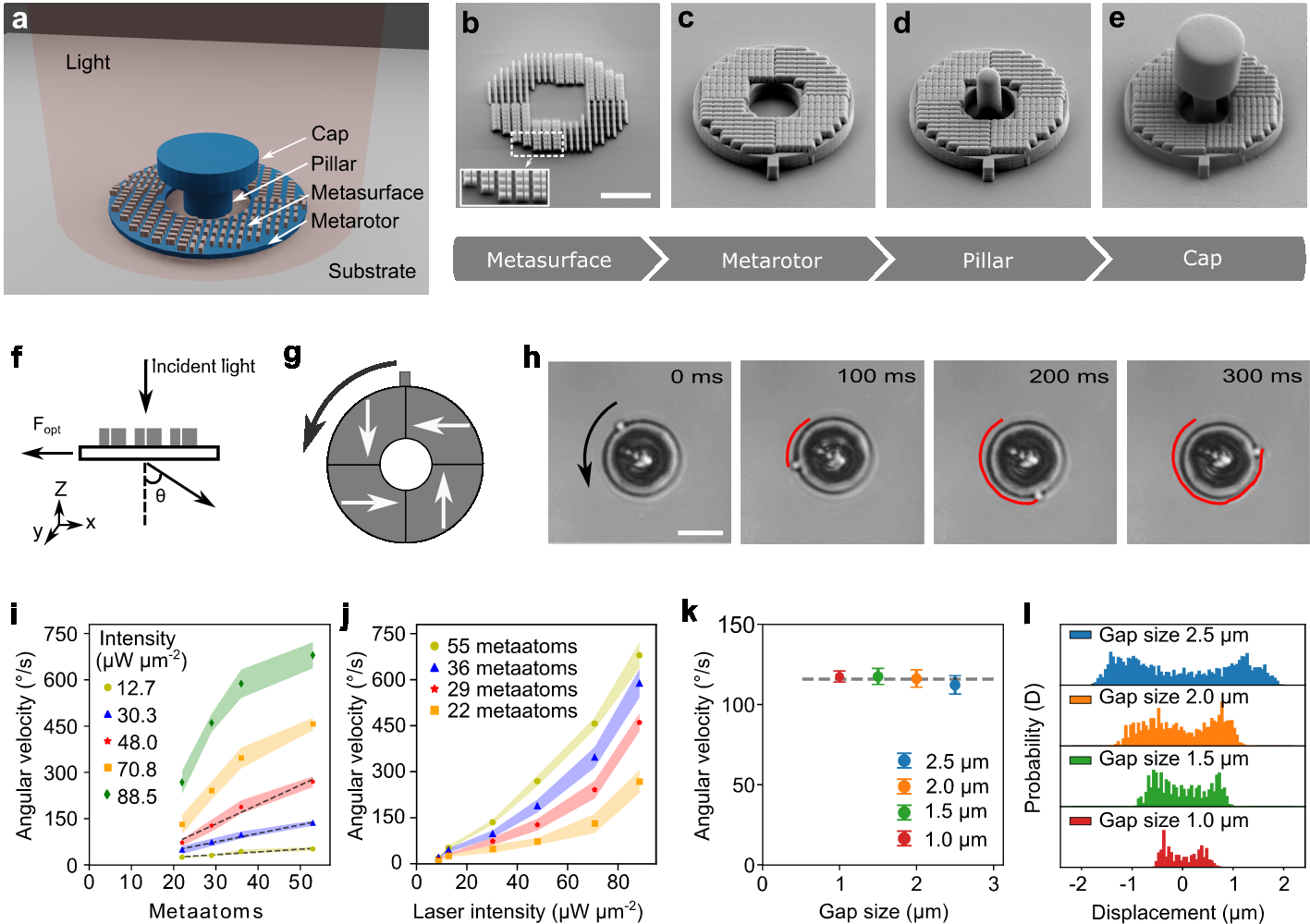}
    \caption{{\bf Metarotors.}
    {\bf a} Schematic illustration of the optically-powered micromotor, featuring a ring-shaped metarotor containing a metasurface, anchored to a glass chip using a capped pillar.
    {\bf b}-{\bf e} Scanning electron microscopy (SEM) images documenting the micromotor fabrication process (scale bar: $5\,{\rm \upmu m}$): 
    {\bf b} A ${\rm Si}$ metasurface is nanofabricated via electron beam lithography; the inset zooms in on the meta-atoms constituting the metasurface.
    {\bf c} A ${\rm SiO_2}$ ring containing the metasurface is etched. 
    {\bf d} The central pillar and {\bf e} the cap are fabricated with SU-8 microlithography.
    {\bf f} Illustration of the deflection of light by the metasurface, resulting in a force acting on the metasurface in the opposite direction.
    {\bf g} The metasurface consists of four segments with different orientations of the meta-atoms. The white arrows indicate the forces they exert onto the metarotor upon illumination with linearly polarized light. The black arrow represents the resulting counterclockwise rotation.
    {\bf h} Optical microscopy images (see Supplementary Video~1) of the rotation of a metarotor under a linearly polarized light beam with an intensity of $35\,{\rm \upmu W \, \upmu m^{-2}}$. The red line indicates the tracking of a protrusion on the metarotor outer border. 
    Scale bar: $10\,{\rm \upmu m}$.
    {\bf i}-{\bf j} Average angular velocities of metarotors with equal diameters ($16\,{\rm \upmu m}$) illuminated by a linearly polarized plane light beam as a function of the number of {\bf i} meta-atoms and {\bf j} laser light intensity.
    {\bf k} Independence of the angular velocity from the gap size between the ring and the pillar.
    {\bf l} Probability distributions of the metarotor position along the x-axis for varying gap sizes. Smaller gap sizes lead to a higher confinement.
    }
    \label{Fig1}
    \end{center}
\end{figure}

We present a rotating micromotor powered by an optical metasurface. This micromotor is constituted by a \emph{metarotor} --- a ring structure containing a metasurface --- that is securely anchored to a glass chip using a capped pillar, as shown in Fig.~\ref{Fig1}a.

The fabrication process involves four key steps. First (Fig.~\ref{Fig1}b), the metasurface is etched, optimized for operation in water under $1064\,{\rm nm}$ plane wave illumination. The metasurface’s unit cells, or \emph{meta-atoms}, are composed of two asymmetric rectangular Si blocks, with dimensions $270\,{\rm nm} \times 200\,{\rm nm} \times 460\,{\rm nm}$ and $400\,{\rm nm} \times 200\,{\rm nm} \times 460\,{\rm nm}$, respectively, separated by a subwavelength gap of $50\,{\rm nm}$ to maximize the efficiency of the $+1$ light diffraction order relative to the $0$ and $-1$ orders \cite{andren2021microscopic}.
Next, a $\rm SiO_2$ ring is etched to support the metasurface (Fig.~\ref{Fig1}c). This is followed by the fabrication of the SU-8 pillar (Fig.~\ref{Fig1}d) and its cap (Fig.~\ref{Fig1}e). The capped pillar anchors the metarotor to the $\rm SiO_2$ substrate, allowing it to rotate freely while suspended in water. The final step involves removing a sacrificial layer between the ring and the substrate.
A comprehensive explanation is provided in the Methods section and Supplementary Fig.~\ref{FigS1}.

This metarotor is powered by the interaction between the incoming light and the metasurface.
By deflecting the incoming light, the metasurface induces a force on the rotating ring in the opposite direction, as schematically shown in Fig.~\ref{Fig1}f. 
By tuning the design of the metasurface, we can control the details of this interaction and therefore the metarotor rotation.

The metasurface of the metarotor in Figs.~\ref{Fig1}a-e  comprises four segments, each with meta-atoms arranged in parallel but rotated by 90° relative to the adjacent segments. The direction of the resulting forces acting on the metarotor under uniform linearly polarized light for each segment are schematically depicted using white arrows in Fig.~\ref{Fig1}g. The collective effect of the four metasurface segments results in a counterclockwise rotation of the metarotor depicted by the black arrow in Fig.~\ref{Fig1}g.
The resulting movement is illustrated in Fig.~\ref{Fig1}h and Supplementary Video~1.

We can control the metarotor's angular velocity either by altering the metasurface design or by modifying the light intensity.
Fig.~\ref{Fig1}i compares metarotors of equal dimensions but varying numbers of meta-atoms under different  light intensities. The angular velocity increases with more meta-atoms (Supplementary Fig.~\ref{FigS2} and Supplementary Video~2). At low intensities ($12.7\, {\rm \upmu W  \,\upmu m^{-2}}$, $30.3\, {\rm \upmu W \, \upmu m^{-2}}$, $48.0\, {\rm \upmu W \, \upmu m^{-2}}$), this increase is linear. However, at higher intensities ($70.8\,{\rm \upmu W \, \upmu m^{-2}}$, $88.5\,{\rm \upmu W \, \upmu m^{-2}}$), the relationship becomes nonlinear due to increased light absorption by the meta-atoms, leading to a local rise in water temperature, reduced viscosity, and decreased rotational viscous drag (Supplementary Fig.~\ref{FigS3}). At low intensities, momentum transfer is the primary factor, while at high intensities, both decreased viscous drag and increased momentum transfer create nonlinear effects (Supplementary Fig.~\ref{FigS4}). Consequently, a nonlinear relationship between the angular velocity and light intensities of different meta-atoms is also observed in Fig.~\ref{Fig1}j (Supplementary Fig.~\ref{FigS5} and Supplementary Video~3).

Finally, we explore the influence of the gap size between the metarotor and the capped pillar, as shown in Fig.~\ref{Fig1}k (and Supplementary Fig.~\ref{FigS6} and in Supplementary Video~4). A minor reduction in angular velocity with an increasing gap size is identified, remaining within the margin of error. Interestingly, the metarotors experience enhanced confinement with a reduced gap size, evident from the probability distribution of the metarotors relative to the center of the pillar (Fig.~\ref{Fig1}l). 

Metarotors can be fabricated in various sizes. We have built a metarotors down to a smallest size of $8\,{\rm \upmu m}$ in diameter, as shown in Supplementary Video~5.

\subsection{Gear trains powered by metarotors}

\begin{figure} [t]
    \begin{center}
    \includegraphics[width=\textwidth]{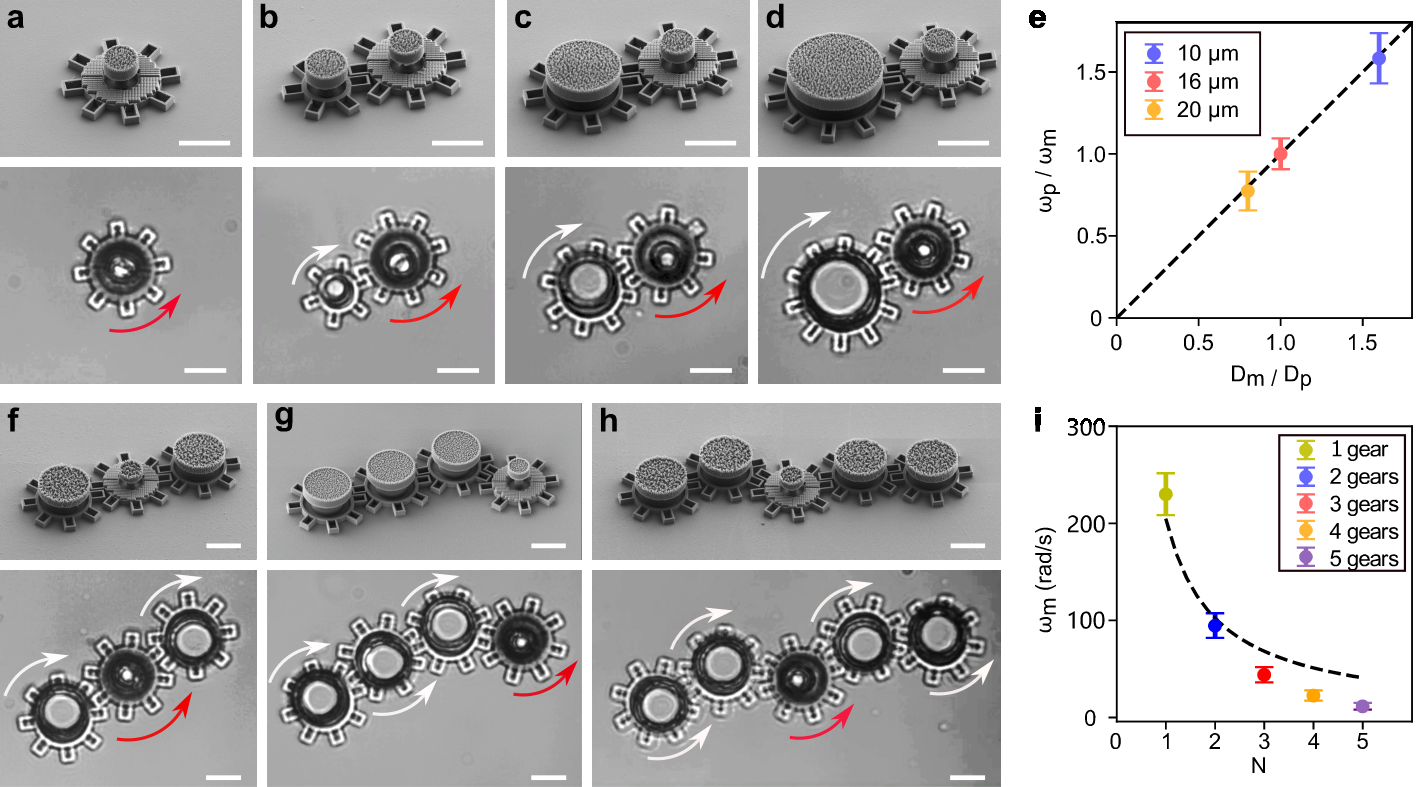}
    \caption{{\bf Gear trains powered by metarotors.} 
    {\bf a}-{\bf d} SEM images (top panels) and optical microscopy images (bottom panels) of metarotors acting as driving gears that propel passive gears with varying diameters ({\bf b}-{\bf d}). The rotation of the metarotors is indicated by the red arrows, while the rotation of the driven gear is indicated by the white arrows.
    {\bf e} The average ratio of angular velocity between the driving gear ($\omega_{\text{m}}$) and the driven gear ($\omega_{\text{p}}$) depends on the ratio of their diameters  
    {\bf f}-{\bf h} SEM images (top panels) and optical microscopy images (bottom panels) of a single driving gear actuating a train of driven gears with the same diameter: {\bf f} $N = 3$, {\bf g} $N = 4$, and {\bf h} $N = 5$ total gears (including the driving gear) with the same diameter $D_{\rm m}=D_{\rm p}$. 
    {\bf i} The angular velocity of the driven gears $\omega = \omega_{\rm m} = \omega_{\rm p}$ versus the number $N$ of gears in the gear train. 
    The dashed line represents $\omega / N$. The optical microscopy images are from Supplementary Video~6.
    Scale bars: $10\,{\rm \upmu m}$.}
    \label{Fig2}
    \end{center}
\end{figure}

Our next step is to employ metarotors to actuate microscopic machines, transferring their work through gear trains.
In the first application, a metarotor serves as a driving gear with a ring diameter $D_{\rm m}$ (a \emph{metagear}, Fig.~\ref{Fig2}a) to propel a driven gear with a ring diameter $D_{\rm p}$ (Figs.~\ref{Fig2}b-d, Supplementary Video~6).
The metagear is fabricated similar to the metarotor described in Figs.~\ref{Fig1}b-e, with the addition of teeth to convert the ring into a gear.
The driven gear is also fabricated like the metarotor but with added teeth and without the metasurface.

As expected, the angular velocity of the passive driven gear, $\omega_{\rm p}$, depends on the angular velocity of the motor metagear, $\omega_{\rm m}$, according to the ratio of their diameters:
\begin{equation}
\omega_{\rm p} = \frac{D_{\rm m}}{D_{\rm p}} \omega_{\rm m}.
\end{equation}
This theoretical relationship is depicted by the dashed line in Fig.~\ref{Fig2}e, with the experimental data represented by their corresponding symbols. 
This system demonstrates the ability to multiply torque when $D_{\rm p} > D_{\rm m}$ or speed when $D_{\rm p} < D_{\rm m}$. 

The metagear can also be used to actuate multiple driven gears at once, as shown in Figs.~ \ref{Fig2}f-h and Supplementary Video~6.
In these examples, all gears have the same diameter and therefore  the same angular velocity $\omega = \omega_{\rm m} = \omega_{\rm p}$.
With the addition of each extra driven gear, $\omega$ decreases. 
We expect this decrease to depend inversely on the number of gears $N$ (i.e., $\omega \propto \frac{1}{N}$, where $N$ is the number of gears) due to increased friction, as shown by the dashed line in Fig.~\ref{Fig2}i. In fact, the experimental results, shown by the symbols in Fig.~\ref{Fig2}i, indicate that $\omega$ decreases faster than expected as $N$ increases, probably because each additional gear introduces new contact surfaces that contribute to higher frictional forces within the system. 

The use of gear trains on the microscale offers the design flexibility needed to create more complex machines.
Examples of gear train configurations with different gear geometries are showcased in Supplementary Video~7.

\subsection{Controlling the metarotor rotation velocity and direction}

\begin{figure} [t]
    \begin{center}
    \includegraphics[width=\textwidth]{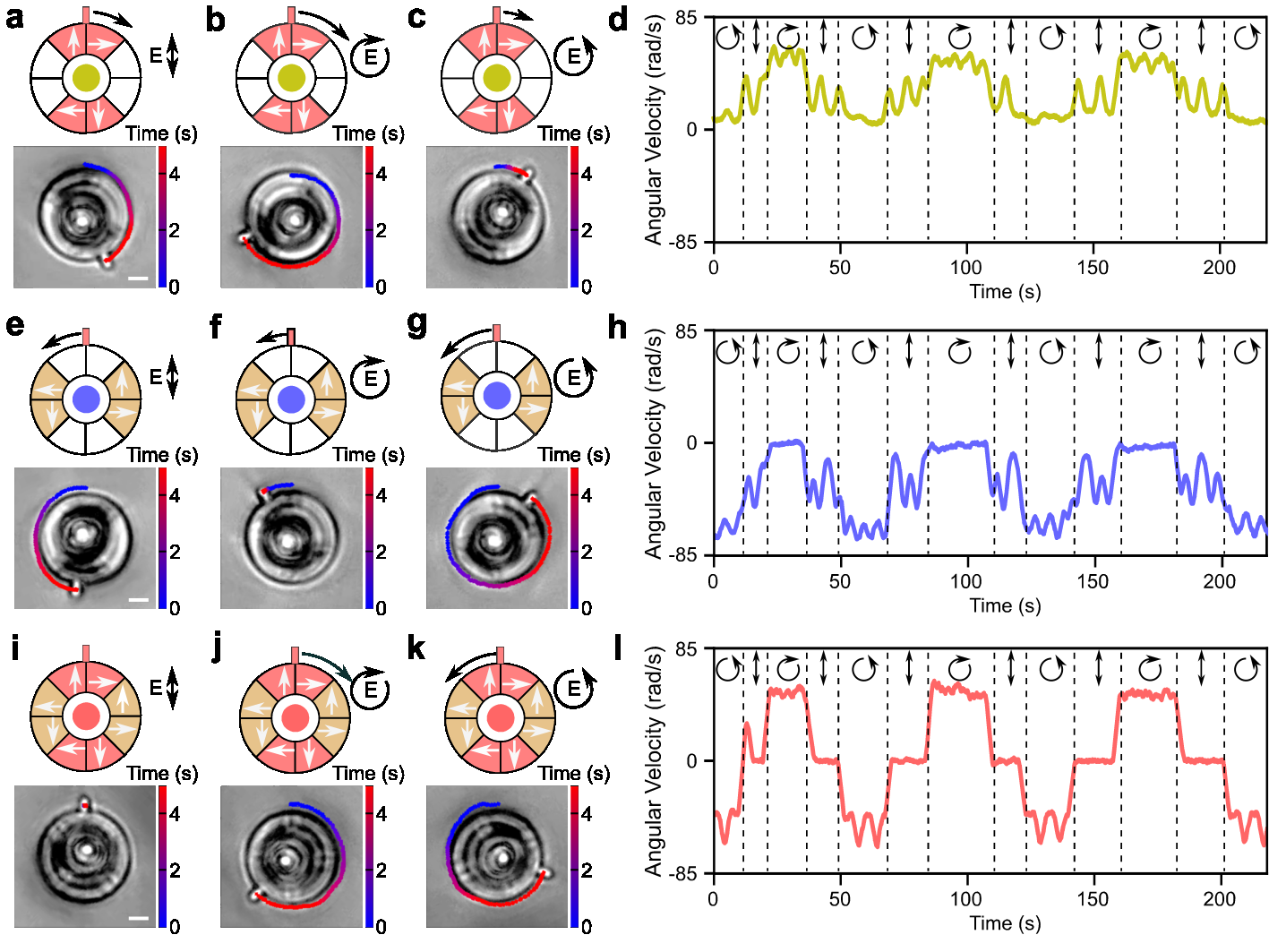}
    \caption{{\bf Metarotor control by metasurface design and light polarization.} {\bf a}-{\bf d} Metarotor with a metasurface design that generates clockwise rotation as a function of light polarization.
    {\bf a}-{\bf c} Schematic illustration of the metasurface design (top) and optical microscopy images (bottom) under illumination with {\bf a} linear polarization, {\bf b} right-hand circular polarization, and {\bf c} left-hand circular polarization. The metasurface segments are indicated by the red-shaded areas and the resulting forces by the corresponding white arrows. The global rotation direction and speed of each motor is indicated by the direction and length of the black arrows.
    The experimentally measured rotation of the metarotor is overlaid on the optical images.
    {\bf d} Angular velocity of the motor with changing light polarization.
    {\bf e}-{\bf h} Metarotor with a metasurface design that generates counterclockwise rotation as a function of the light polarization. 
    {\bf i}-{\bf j} Metarotor with a metasurface design combining the designs in ({\bf a}-{\bf c}) and ({\bf e}-{\bf g}) as a function of the light polarization. The metarotor remains stationary under linear polarization, rotates counterclockwise under left-hand circular polarization, and clockwise under right-hand circular polarization. The optical microscopy images are taken from Supplementary Video~8.
    Scale bars: $5\,{\rm \upmu m}$.
    }
    \label{Fig3}
    \end{center}
\end{figure}

Until now, we have used only linearly polarized light to power the metarotors. Remarkably, by altering the polarization of the illumination, we can also dynamically control their rotational direction.

We first examine a metarotor with metasurface segments generating clockwise forces (Figs.~\ref{Fig3}a-d). The metarotor ring is divided into eight segments with only four containing metasurfaces, as shown by the shaded areas in the schematics on the top of Figs.~\ref{Fig3}a-c;
the white arrows indicate the forces generated by each segment under the various illumination conditions, while the black arrow represents the overall metarotor rotation direction (which is always clockwise for this metasurface design) and magnitude (which varies for different illumination polarizations).
A measured 5-second trajectory is overlaid on the optical images below each schematic in Figs.~\ref{Fig3}a-c.
Under linearly polarized light, this metarotor rotates clockwise (Fig.~\ref{Fig3}a). A right-hand circular polarization increases the metarotor angular velocity (Fig.~\ref{Fig3}b), while a left-hand circular polarization decreases it (Fig.~\ref{Fig3}c). This variation is due to the superposition of the change in momentum as the incident light is deflected by the metasurface and of spin angular momentum (SAM) transferred to the metarotor \cite{allen1992orbital, he1995direct}. 

Next, we consider a specularly symmetric design of the metasurfaces, as shown in the schematics in Figs.~\ref{Fig3}e-g.
This design results in opposite forces and therefore a counterclockwise rotation.
For linear polarization (Fig.~\ref{Fig3}e), the metarotor exhibits a rotational velocity similar to that in the previous case  (Fig.~\ref{Fig3}a) but in the opposite direction. 
Under circular polarization (Figs.~\ref{Fig3}f,g), the rotational velocity decreases for right-handed polarization and increases for left-handed polarization. 
This is the opposite behavior to that observed in the previous case (Figs.~\ref{Fig3}b,c), which can also be observed when varying the illumination polarization (Fig.~\ref{Fig3}h).
This is due to the fact that, while the metasurface light deflection is the opposite than in the previous case, the transfer of SAM to the metarotor is the same as in the previous case for each polarization.

In order to build a metarotor that can rotate in either direction depending on polarization, we can now combine these two designs as shown in Figs.~\ref{Fig3}i-k.
Now, under linear polarization, the forces cancel out, resulting in no movement (Fig.~\ref{Fig3}i). 
Under right-hand circular polarization, the metarotor rotates clockwise (Fig.~\ref{Fig3}j), and, under left-hand circular polarization, it rotates counterclockwise (Fig.~\ref{Fig3}k).
This permits to dynamically adjust the rotation direction and velocity of the metarotor by the illumination polarization (Fig.~\ref{Fig3}l).

Conveniently, these designs can be fabricated on the same chip, enabling distinct behaviors by changing the light polarization (Supplementary Video~8). This capability extends to altering the rotational direction within gear trains as well (Supplementary Video~9).

\subsection{Responsive metamachines transferring rotational to linear movement}

\begin{figure} [t]
    \begin{center}
    \includegraphics[width=\textwidth]{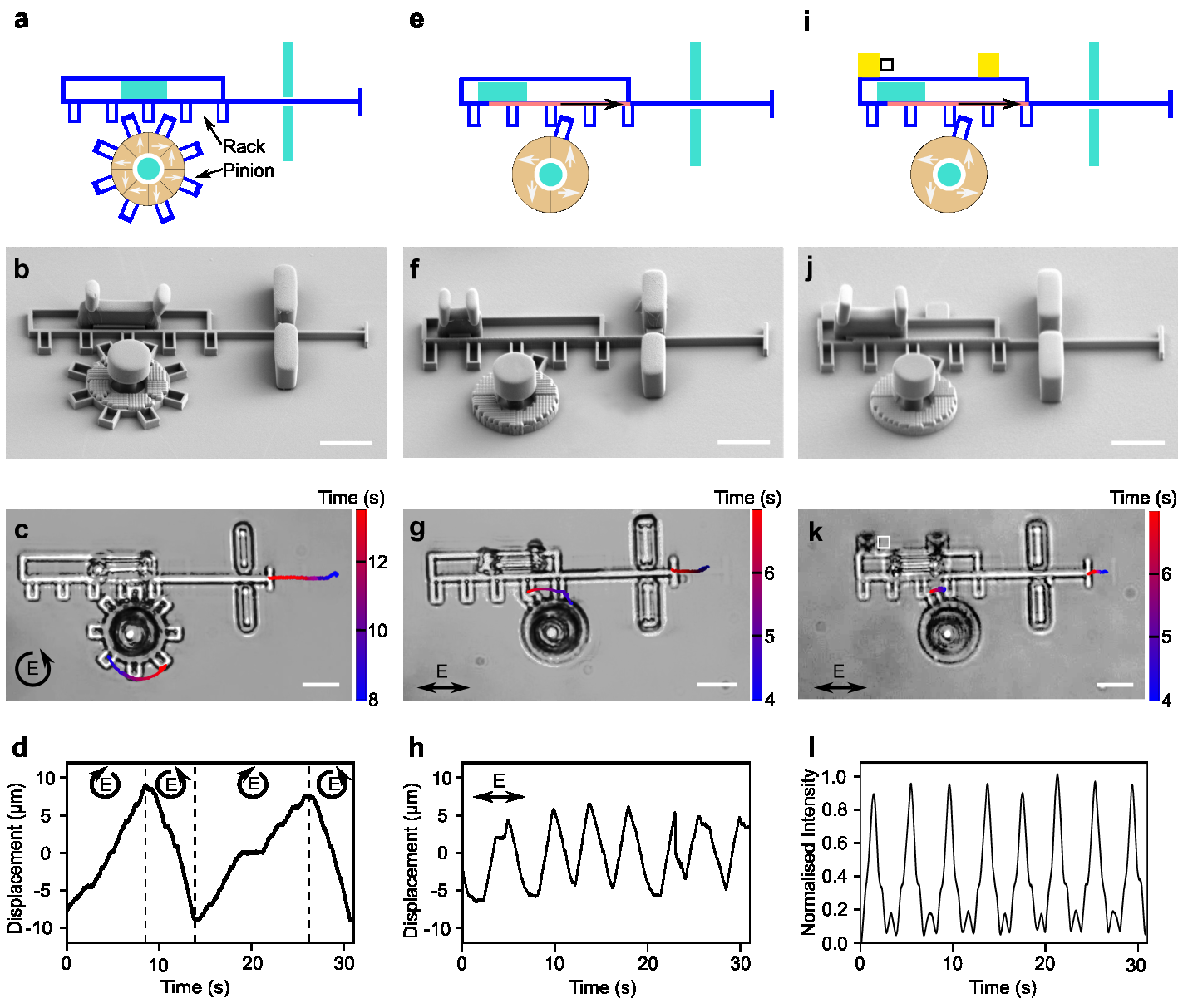}
    \caption{{\bf Microscopic rack and pinion metamachines.} 
    {\bf a}, {\bf e}, {\bf i} Schematic illustrations of three designs of rack and pinion metamachines to convert the rotational motion produced by a metarotor into linear motion.
    The movable rack and pinion are shown in blue, while the immobile parts are in turquoise.
    Metasurface segments are highlighted in red and yellow. The forces generated on the driving gear and rack under illumination are depicted by white and black arrows, respectively. 
    {\bf b}, {\bf f}, {\bf j} Corresponding SEM images, and {\bf c}, {\bf g}, {\bf k} optical microscopy images from Supplementary Video~10, 12, 13.
    {\bf d} The pinion metarotor is equipped with a metasurface designed so that its rotation direction is different for left and right-handed circularly polarized light, enabling forward motion of the rack under right-handed circular polarization and backward motion under left-handed circular polarization, allowing dynamic back-and-forth motion by changing the circular polarization of light. 
    {\bf h} Equipping both the rack and the pinion with metasurfaces permits oscillatory rack movement under constant linearly polarized light. The pinion has a single tooth that periodically moves the rack leftward when engaged, while the rack's metasurface moves it rightward, mimicking a macroscopic spring. Balancing the forces of both metasurfaces achieves oscillatory rack movement under linearly polarized light. 
    {\bf l} The same rack and pinion design can periodically move a gold mirror (illustrated in golden yellow in {\bf i}), changing the average amount of transmitted light {\bf l} through the white box shown in {\bf i} and {\bf k}. 
    Scale bars: $10\,{\rm \upmu m}$.
    }
    \label{Fig4}
    \end{center}
\end{figure}

Having developed metarotors capable of dynamically altering their rotation direction, we use them to build some microscopic machines with movable parts also to convert rotational motion into linear motion in a rack and pinion design (Fig.~\ref{Fig4}). 
In this design (Figs.~\ref{Fig4}a-c), the metarotor acts as the pinion and engages a rack.
When exposed to right-handed circularly polarized light, the metarotor rotates clockwise, moving the rack to the right. 
Conversely, under left-handed circularly polarized light, the metarotor rotation direction is reversed, causing the rack to move to the left. By alternating these two polarizaitons, we can obtain an oscillatory back-and-forth motion of the rack, as shown in Fig.~\ref{Fig4}d. The amplitude and frequency of this motion can be adjusted by controlling the light intensity and the intervals between changes in polarization. 

Next, we can consider an alternative design (Figs.~\ref{Fig4}e-g) that will permit us to oscillate the rack under constant linearly polarized light.
To achieve this, we add a metasurface to the rack, as shown by the red-shaded area in Fig.~\ref{Fig4}e.
The metarotor, now featuring a single tooth, moves the rack leftward when engaged, while the rack metasurface pushes it rightward, mimicking the behavior of a microscopic spring. A delicate balance between these forces, which can be optimized by the number of meta-atoms (Supplementary Fig.~\ref{FigS7}, Supplementary Video~11, Supplementary Video~12), results in an oscillatory back-and-forth movement of the rack under constant linearly polarized light (Fig.~\ref{Fig4}h). While this design offers comparatively lower tunability and flexibility compared to the previous one, its advantage lies in obtaining an oscillatory rack motion without the need to repeatedly alter the polarization of the light.

Finally, we demonstrate how this design can be used to control the movement of some microscopic gold mirrors and therefore the reflection and transmission of light on the microscopic scale (Figs.~\ref{Fig4}i-l).
By embedding gold mirrors in the rack, indicated by the golden yellow areas in the schematic in Fig.~\ref{Fig4}i, we can dynamically block and deflect light within an optical device (Fig.~\ref{Fig4}l, Supplementary Video~12). 

In comparison with previously reported field-driven micromotors~\cite{kim2014ultrahigh,shields2018supercolloidal,hong2024optoelectronically, shi2024dna,xia2010ferrofluids,zandrini2017magnetically,wang2017dynamic,wang2021light,liu2022creating,friese2001optically,kelemen2006integrated, bianchi2018optical,aubret2018targeted}, this highlights the advantage of adopting a design and fabrication process compatible with metal-oxide-semiconductor (CMOS) technology, as it permits us to ensures a straightforward implementation with adherence to reproduction standards in cleanroom facilities as well as seamless parallelization of micromotors on a single chip. Additionally, it enables the direct integration of various other CMOS-based components, further amplifying the versatility and efficiency of our advanced manufacturing approach.

\subsection{Conclusion}

In summary, our work introduces optically powered micromotors capable of performing work under broad light illumination, utilizing integrated metasurfaces. This innovation achieves motor miniaturization to diameters below $10\,{\rm \upmu m}$, enabling straightforward parallelization and precise control over individual micromotor units. Control over rotation direction and speed can be achieved through metasurface design or adjustments in light intensity and polarization.

Furthermore, these micromotors can be assembled into functional microscopic metamachines, transducing rotational motion in gear trains or converting it to linear motion using rack and pinion designs. The on-chip fabrication process, compatible with standard CMOS lithography, facilitates seamless integration with other CMOS components like metalenses and plasmonic sensors. Future developments may involve arrays of micromotors for collective manipulation of objects and flow control on the micron scale. The use of structured light could further enhance flexibility and tunability, allowing precise control over individual micromotors.

By using light as a widely available and biocompatible energy source, these micromotors are particularly suited for manipulating biological matter, including bacteria and cells. This non-toxic energy source broadens the scientific and technological applications of our light-driven micromotors and metamachines.

Looking ahead, we anticipate integrating our metamachines with planar optical elements, such as high-numerical-aperture metalenses. These metalenses can focus incoming light, enabling precise manipulation of colloidal objects. This synergy could lead to advanced applications, such as mechanically programmable devices that alter optical properties through planar optical elements, including the generation of spatially structured lights. Our metasurfaces can elevate optical force modulation to the femtonewton (fN) scale, making them valuable on-chip force measuring instruments for assessing mechanical properties of individual cells or biological macromolecules, such as DNA.

\section{Methods}

\subsection {Metarotor fabrication protocol}

The fabrication of optically powered microscopic motors and micromachines relies on cleanroom fabrication methods, which can be divided into four steps: preparation of the substrate, fabrication of the metasurface, etching of the metarotor bearing the metasurface, and fabrication of the pillar and of the cap to anchor the ring. 
The details are schematically illustrated in Supplementary Fig.~\ref{FigS1}.

\subsubsection{Preparation of the substrate}

For the fabrication, we used a 4-inch fused silica wafer. An $800\,{\rm nm}$ layer of amorphous silicon (a-Si) was deposited using low-pressure chemical vapor deposition (LPCVD, Furnace Centrotherm) to serve as a sacrificial layer, facilitating the later release of the ring from the substrate. Subsequently, a $400\,{\rm nm}$ layer of ${\rm SiO_2}$ was deposited via plasma-enhanced chemical vapor deposition (PECVD, Oxford PlasmaPro 100), serving as the material for crafting the metarotor. Finally, a $460\,{\rm nm}$ a-Si layer was deposited using LPCVD (Furnace Centrotherm), to be etched to fabricate the metasurface.

\subsubsection{Fabrication of the metasurface}

To pattern the a-Si for metasurface fabrication, a $300\,{\rm nm}$ thick layer of APR6200 (CSAR ARP 6200.13) resist was spin-coated on the substrate. Subsequently, electron beam lithography (Raith EBPG 5200) was employed for exposure and the resist was developed using a developer (n-Amyl acetate). Next, a metal layer composed of $2\,{\rm nm}$ chromium (Cr) and $40\,{\rm nm}$ nickel (Ni) was evaporated (Kurt J. Lesker PVD225) onto the patterned resist. This was followed by the subsequent lift-off process in an organic remover (Microposit Remover 1165). The metal layer served as a hard mask for undergoing the chlorine (${\rm Cl_2}$) reactive ion etching process (Oxford Plasmalab 100), enabling the precise etching of a $460\,{\rm nm}$ a-Si metasurface. After completing this process, the metallic layer was removed using a chemical etching method (SunChem Nickel/Chromium etchant). Finally, a $550\,{\rm nm}$ ${\rm SiO_2}$ layer was deposited onto the metasurface using PECVD (Oxford plasmaPro 100) to serve as a protection layer.

\subsubsection{Fabrication of the metarotor and movable parts}

For the metarotor and the movable parts of the geared micromachines, two patterning methods were used depending on the required precision. For high-precision parts, EBL (Raith EBPG 5200) was used to define the pattern on the resist ARP6200 (CSAR ARP 6200.13). A metal layer consisting of $2\,{\rm nm}$ Cr and $60\,{\rm nm}$ Ni was evaporated (Kurt J. Lesker PVD225) onto a patterned resist, followed by the subsequent lift-off process in Remover-1165 (Microposit Remover 1165). For parts tolerating lower precision, direct laser writing (Heidelberg, MLA 150) was employed to define a double-layer photoresist, LOR3A/S1805. LOR3A (MicroChem Photoresist) acted as a sacrificial release layer for the lift-off process, while S1805 (Shipley Photoresist) served as the top positive resist. The exposed resist was developed using a developer (Microposit MF-CD 26) and then employed for lifting off the deposited $2\,{\rm nm}$ Cr and $60\,{\rm nm}$ Ni as a durable hard mask for the subsequent etching step. In both cases, a reactive ion etching process (Oxford Plasmalab 100) with a mixture of fluoroform (${\rm CHF_3}$) and argon (Ar) gases was employed to etch the ${\rm SiO_2}$ layer. Following this, ${\rm Cl_2}$ reactive ion etching (Oxford Plasmalab $100$) was applied to remove the exposed a-Si between the PECVD-deposited ${\rm SiO_2}$ and the fused silica substrate. 

\subsubsection{Fabrication of the pillar, cap, and immobile parts}

The immobile components, including the pillars and caps employed to anchor the micromotors and micromachine parts, were produced through a direct laser writing process. Initially, a $3.8\,{\rm \upmu m}$ layer of SU8-3005 (MicroChem Photoresist), a negative photoresist, was spin-coated onto the substrate. Direct laser writing (Heidelberg, MLA150) was then utilized to expose the photoresist, defining pillars in the center of the rotating rings and micromachine parts. After developing the SU-8 3005 in Remover500 (Microresist Technology), the sample underwent a 10-minute hard bake at $160\,^{\circ}{\rm C}$  to enhance the durability of the pillars.

Subsequently, a $4\,{\rm \upmu m}$ layer of positive photoresist AZ4533 (Clariant Photoresist) was spin-coated onto the substrate as a sacrificial layer to create caps on the pillars. A $250\,{\rm W}$ ${\rm O_2}$ plasma (Plasma-Therm) was applied for 1 minute to ensure that the upper surface of the SU-8 pillar remained uncovered by any AZ4533 residues, ensuring successful linking of the pillar and cap. The cap was fabricated by spin-coating a $3.8\,{\rm \upmu m}$ SU-8-3005 layer on top of the sacrificial AZ4533 layer, which was then defined using direct laser writing. SU-8 3005 was initially developed in Remover500, after which AZ4533 was removed by immersing the sample in acetone (Sigma Aldrich) to separate the cap from the disks. In the final step, the rotating rings and micromachine parts were released from the substrate by removing the sacrificial a-Si layer using highly selective xenon(II) fluoride (${\rm XeF_2}$) ion etching.

\subsubsection {Fabrication of gold mirrors}

The fabrication of the gold mirrors took place before depositing $\rm SiO_2$ on the metasurface. The structure of the mirrors was defined using direct laser writing (Heidelberg, MLA $150$) on the substrate coated with a double layer of photoresist LOR3A/S$1805$, achieved through spin-coating. The exposed resist was developed using a developer (MF-CD $26$), followed by the deposition of a $60\,{\rm nm}$ gold layer onto the resist (Kurt J. Lesker PVD$225$). Subsequently, the gold mirrors were formed by conducting a lift-off process to remove the unexposed photoresist in Remover 1165. The subsequent fabrication processes were then carried out as previously described.

\subsection{Scanning electron microscopy imaging}

For scanning electron microscopy (SEM) imaging, the samples underwent sputter-coating with a $15\,{\rm nm}$ gold layer. SEM images were captured using a Zeiss Supra $55$, operating at a current of $5\,{\rm kV}$, and an SE2 detector.

\subsection{Optical setup}

A schematic of the optical setup is shown in Supplementary Fig.~\ref{FigS8}. The optical setup consists of two parts: the bright--filed home--built optical microscope to observe the motion of micromotors and micromachines; and the illumination part based on a  $1064\,{\rm nm}$ laser to make the micromotors move. The incident laser beam is weakly focused from the upper surface of the chip using a lens, generating a beam with an approximate waist diameter of $300\,{\rm \upmu m}$ on the chip. By using half-wave and quarter-wave plates, we controlled the incident light polarization, while the power of the laser on the sample is regulated manually using the combination of a half-wave plate and a polarizing beam splitter after the laser. The videos were acquired by using a CCD camera.

\subsection{Measurements}

The measurements were conducted by creating a $4\,{\rm \upmu L}$ sample cell, which consisted of the chip as bottom slide, a $120\,{\rm \upmu m}$ thick polydimethylsiloxane (PDMS) spacer, and a coverslip. The cell was filled with an aqueous solution containing $0.005\, {\rm wt\%}$ of Triton X-$100$ (Sigma Aldrich) to prevent the spinning disks from sticking to the substrate. Additionally, the chip was briefly inserted into an ultrasonic bath for $10\,{\rm s}$ to ensure none of the motors were stuck. The positions and orientations of microrotors or micromachines were examined through a custom-written Python code.

\subsection{Photothermal heating simulation}

The temperature profile around the micromotor is evaluated by modeling the a-Si meta-atoms as the exclusive sources of photothermal heating, incorporating a complex refractive index $n = 3.8 + i0.0064$ as determine from ellipsometry. The heat source density $Q$ is derived through simulations of absorption cross-sections via field element calculations in COMSOL. The model include both heat conduction and convection, with room tempearture ($293\,{\rm K}$) set as the boundary condition. For a-Si, a thermal conductivity is $1.8\, \mathrm{W/m \cdot K}$ and density is $2.329 \, \mathrm{kg\,m^{-3}}$.

\section{Acknowledgments}

The authors would like to acknowledge funding from the H2020 European Research Council (ERC) Starting Grant ComplexSwimmers (Grant No. 677511) (GV), the Horizon Europe ERC Consolidator Grant MAPEI (Grant No. 101001267) (GV), the Marie Sklodowska-Curie Individual Fellowship (Grant No.101064381) (MR), and the Knut and Alice Wallenberg Foundation (Grant No. 2019.0079) (GV). Fabrication in this work was done at Myfab Chalmers. 

\section{Contributions}

G.W. and G.V conceived the idea, G.W. developed the fabrication process, performed the experiments, and analyzed the data. M.R. and K.L. assisted with fabrication and measurements. G.P., A.C., and G.W. built the optical setup. M.S. conducted the thermal absorption simulation. G.V. and M.K. supervised the study. G.W., M.R., and G.V. wrote the manuscript, with contributions from all other coauthors.

\section*{Competing interests}

The authors declare that there are no competing financial or non-financial interests. 

\bibliographystyle{naturemag}
\bibliography{references.bib}

\newpage

\section{Supplementary Information}
\begin{suppfigure}[h!]
    \centering
    \includegraphics[width=\textwidth]{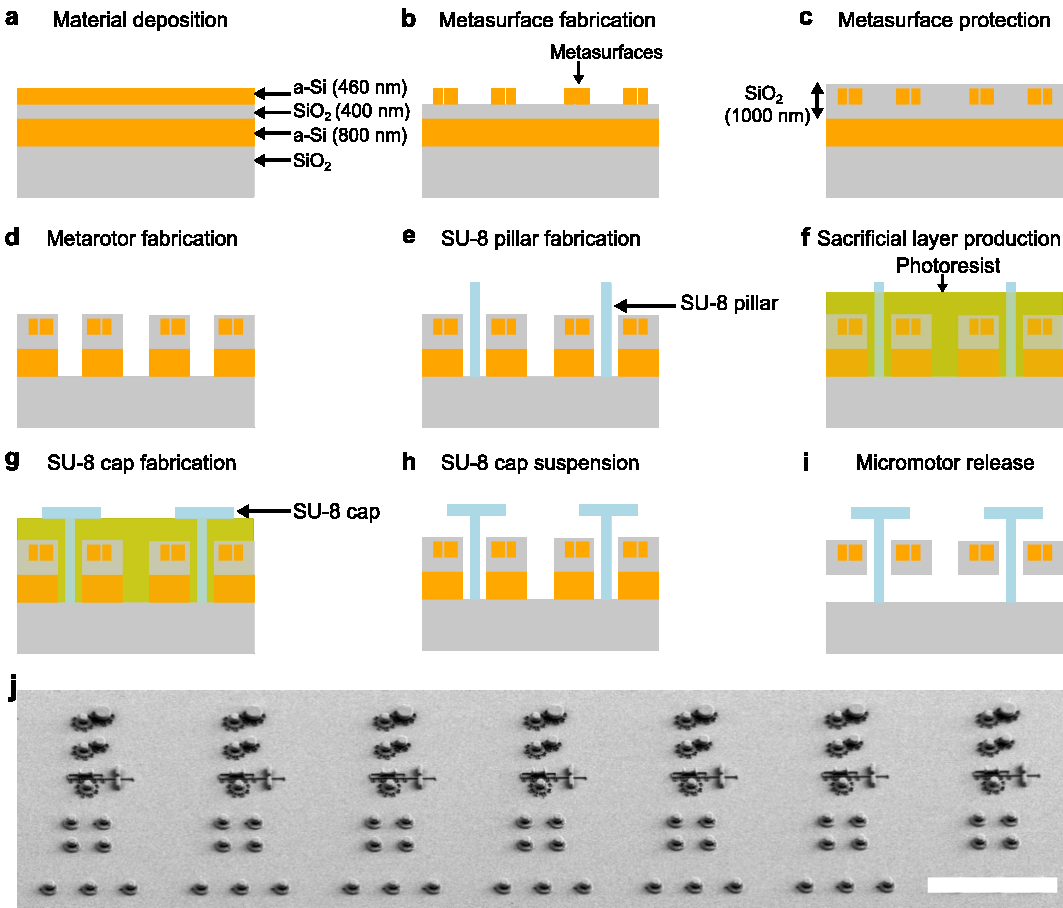}
    \caption{{\bf Fabrication process of the metarotors}. 
    {\bf a} The substrate is a 4-inch fused silica wafer with deposited layers of $800\,{\rm nm}$ amorphous silicon (a-Si), $400\,{\rm nm}$ ${\rm SiO_2}$, and $460\,{\rm nm}$ a-Si.
    {\bf b} The metasurface is fabricated through electron beam lithography (EBL) and subsequent reactive ion etching (RIE) of the $460\,{\rm nm}$ a-Si layer.
    {\bf c} A $600\,{\rm nm}$ ${\rm SiO_2}$ layer is deposited to encapsulate and protect the metasurface.
    {\bf d} The metarotor is fabricated using EBL followed by RIE of the $1000\,{\rm nm}$ ${\rm SiO_2}$ and $800\,{\rm nm}$ a-Si layers.
    {\bf e} The SU-8 pillar is fabricated using direct laser writing.
    {\bf f} A sacrificial positive photoresist layer is spin-coated with its thickness deliberately kept slightly below the height of the pillar.
    {\bf g} The SU-8 cap is fabricated using direct laser writing.
    {\bf h} The sacrificial layer is released in acetone.
    {\bf i} The metarotor is released through selective etching of the sacrificial a-Si layer.
    {\bf j} A low-magnification scanning electron microscope (SEM) image demonstrates the parallelization of micromotor and metamachine fabrication. Scale bar: $100\,{\rm \upmu m}$.
    }
    \label{FigS1}
\end{suppfigure}

\begin{suppfigure}[p]
    \centering
    \includegraphics[width=\textwidth]{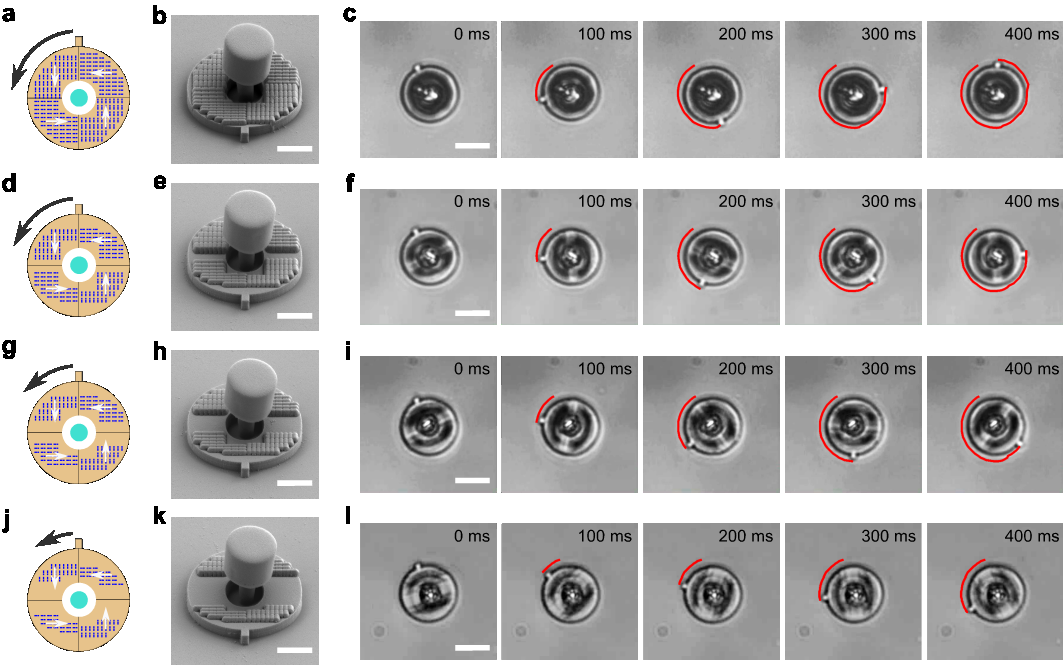}
    \caption{{\bf Angular velocity of metarotors vs number of meta-atoms within their metasurface.}
    {\bf a}, {\bf d}, {\bf g}, {\bf j} Illustrations of metarotor designs with identical dimensions but varying numbers of meta-atoms: {\bf a} 55, {\bf d} 36, {\bf g} 29, and {\bf j} 22 meta-atoms in each of the four surface segments. The dark blue rectangles represent the meta-atoms. The white arrows indicate the forces they exert onto the rotating disk under linearly polarized light. The cyan circle in the center illustrates the immobile pillar. The black arrow illustrates the rotation direction and speed under linearly polarized light. 
    {\bf b}, {\bf e}, {\bf h}, {\bf k} Corresponding scanning electron microscope (SEM) images. Scale bars: $5\,{\rm \upmu m}$. 
    {\bf c}, {\bf f}, {\bf i}, {\bf l} Optical microscopy images of the metrotors rotating under linearly polarized light at different time intervals: $0\,{\rm ms}$, $100\,{\rm ms}$, $200\,{\rm ms}$, $300\,{\rm ms}$, and $400\,{\rm ms}$. Scale bars: $10\,{\rm \upmu m}$. The red line tracks the rotation of the micromotors. A larger number of meta-atoms within each metasurface generates a faster rotation of the micromotors.
    }
    \label{FigS2}
\end{suppfigure}

\begin{suppfigure}[p]
    \centering
    \includegraphics[width=\textwidth]{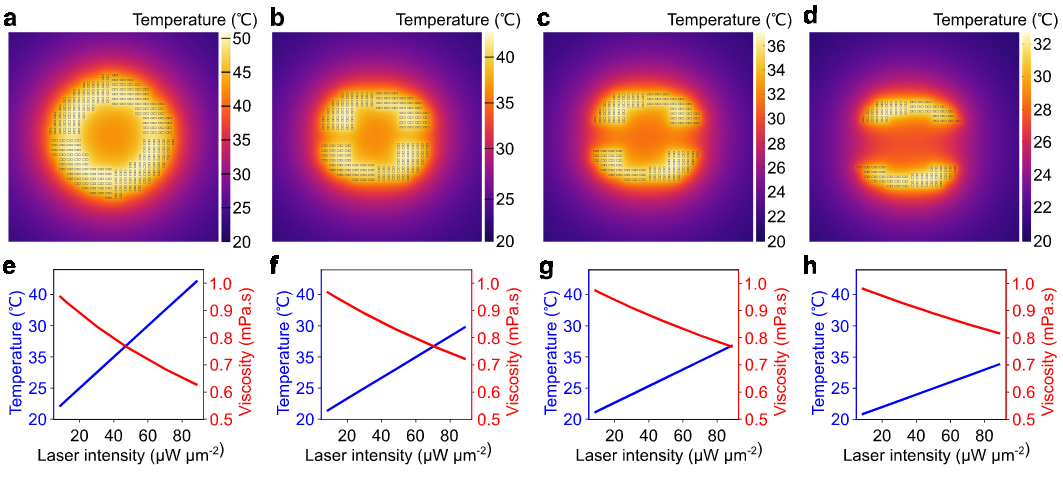}
    \caption{{\bf Finite element simulation of temperature distribution in four motors with varying amounts of meta-atoms}. {\bf a}-{\bf d} Absolute temperature distribution, obtained from finite element simulation, around a motor with {\bf a} 55, {\bf b} 36, {\bf c} 29, and {\bf d} 22 meta-atoms in each metasurface section illuminated with a $120\,{\rm \upmu W \, \upmu m^{-2}}$ s-polarized incident plane wave ($\lambda = 1064\,{\rm \rm nm}$). {\bf e}-{\bf h} Simulated temperature around a motor with {\bf e} 55, {\bf f} 36, {\bf g} 29, {\bf h} 22 meta-atoms, alongside calculated dynamic viscosity of water, as a function of incident light intensity. 
    }
    \label{FigS3}
\end{suppfigure}

\begin{suppfigure}[p]
    \centering
    \includegraphics[width=\textwidth]{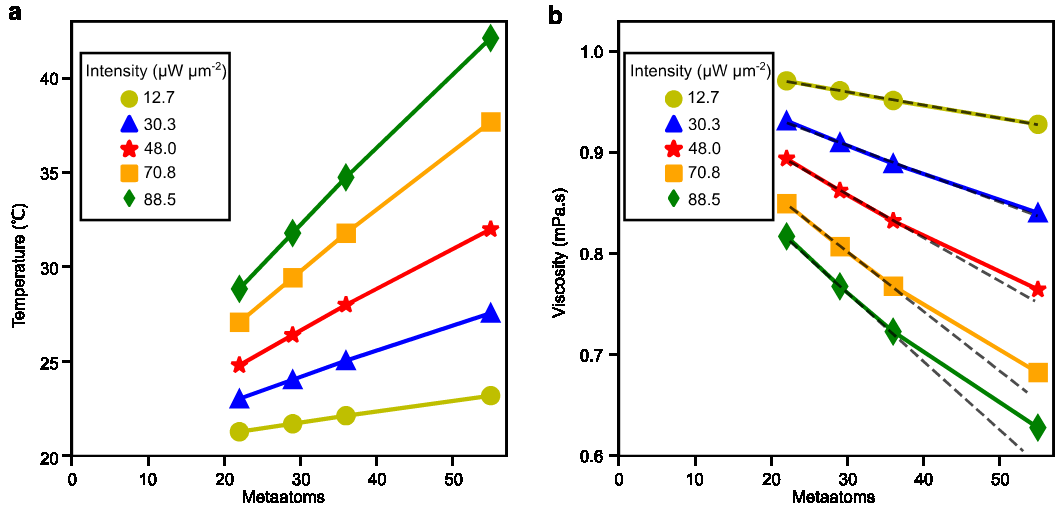}
    \caption{{\bf Temperature and viscosity of four metarotors with varying numbers of meta-atoms at different light intensities}. {\bf a} Simulated temperature of four motors with 22, 29, 36, and 55 meta-atoms in each metasurface section at five light intensities: $12.7\,{\rm \upmu W \, \upmu m^{-2}}$, $30.3\,{\rm \upmu W \, \upmu m^{-2}}$, $48.0\,{\rm \upmu W \, \upmu m^{-2}}$, $70.8\,{\rm \upmu W \, \upmu m^{-2}}$, and $88.5\,{\rm \upmu W \, \upmu m^{-2}}$. The temperature increases almost linearly with the number of meta-atoms at all intensities. 
    {\bf b} Calculated viscosities for four metarotors with 22, 29, 36, and 55 meta-atoms in each metasurface section at the same five light intensities. At low intensities ($12.7\,{\rm \upmu W \, \upmu m^{-2}}$, $30.3\,{\rm \upmu W \, \upmu m^{-2}}$, $48.0\,{\rm \upmu W \, \upmu m^{-2}}$), the relationship between viscosity and number of meta-atoms is almost linear. At higher intensities ($70.8\,{\rm \upmu W \, \upmu m^{-2}}$, $88.5\,{\rm \upmu W \, \upmu m^{-2}}$), viscosities descrease less than linearly with more meta-atoms.  
    }
    \label{FigS4}
\end{suppfigure}

\begin{suppfigure}[p]
    \centering
    \includegraphics[width=\textwidth]{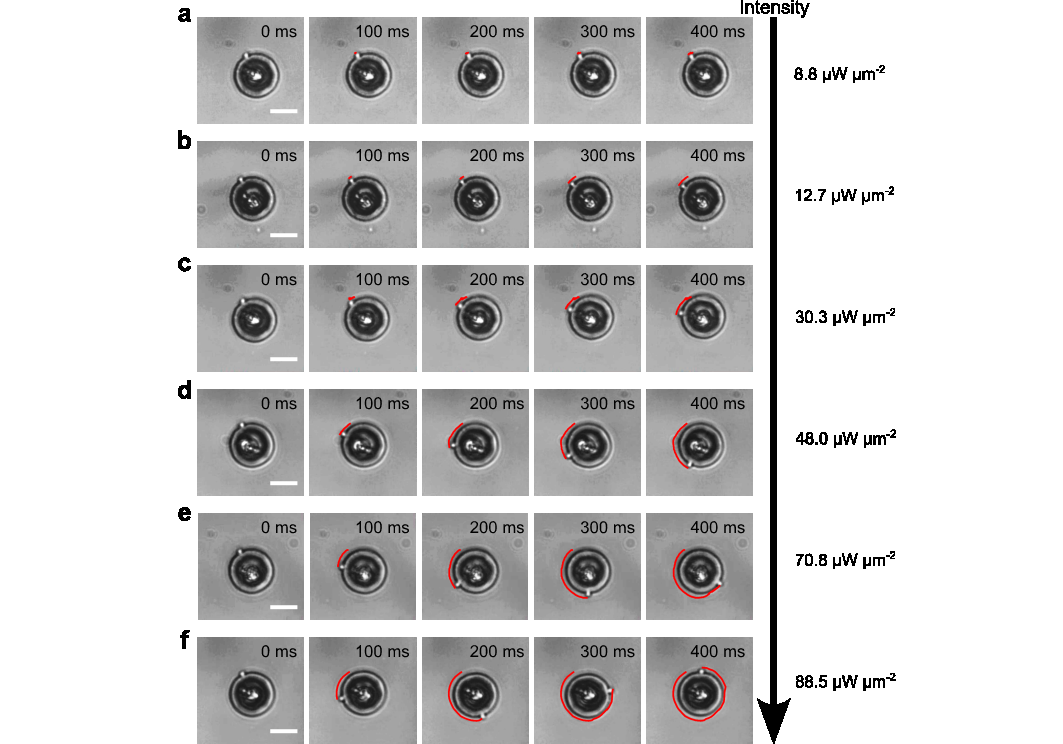}
    \caption{{\bf Angular velocity of metarotors vs light intensity}. 
    Optical microscopy images of the rotation of metarotors under linearly polarized light with different intensities: {\bf a} $8.8\,{\rm \upmu W \, \upmu m^{-2}}$, {\bf b} $12.7\,{\rm \upmu W \, \upmu m^{-2}}$, {\bf c} $30.3\,{\rm \upmu W \, \upmu m^{-2}}$, {\bf d} $48.0\,{\rm \upmu W \, \upmu m^{-2}}$, {\bf e} $70.8\,{\rm \upmu W \, \upmu m^{-2}}$, and {\bf f} $88.5\,{\rm \upmu W \, \upmu m^{-2}}$. The images were captured at times $0\,{\rm ms}$, $100\,{\rm ms}$, $200\,{\rm ms}$, $300\,{\rm ms}$, and $400\,{\rm ms}$. The micromotors rotate faster as the light intensity increases. 
    Scale bars: $10 \,{\rm \upmu m}$. 
    }
    \label{FigS5}
\end{suppfigure}

\begin{suppfigure}[p]
    \centering
    \includegraphics[width=\textwidth]{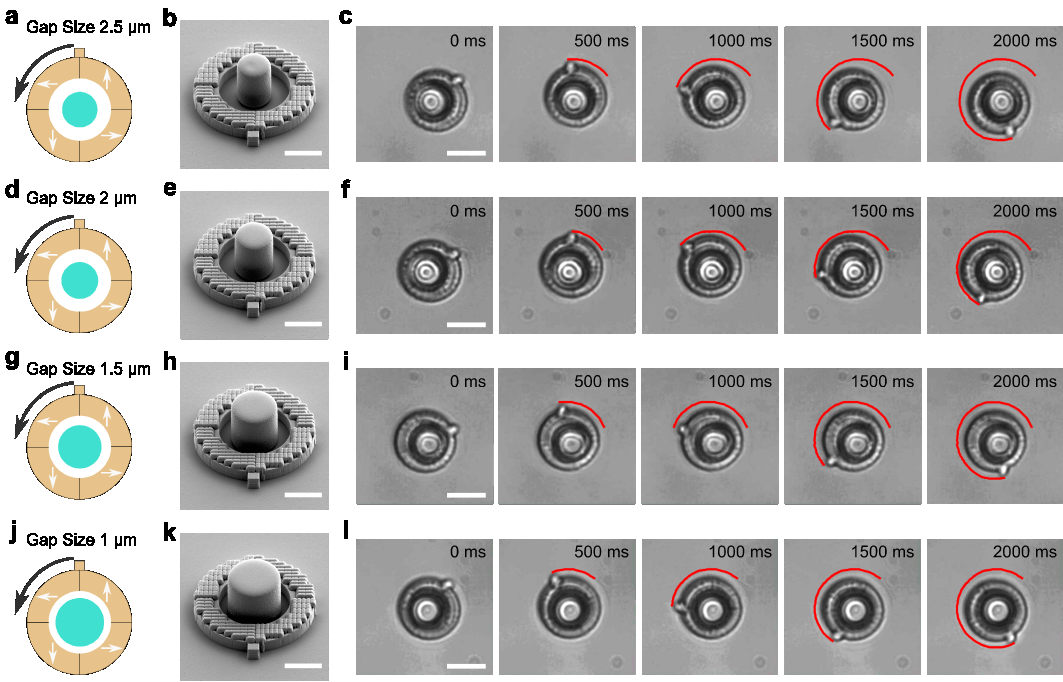}
    \caption{{\bf Metarotor rotation as a function of gap size between the rotating ring and the immobile pillar}. 
    {\bf a}, {\bf d}, {\bf g}, {\bf j} Illustrations of the metarotors featuring an identical ring and metasurface design but varying gap sizes between the rotating ring and the central pillar: {\bf a} $2.5 \,{\rm \upmu m}$, {\bf d} $2\,{\rm \upmu m}$, {\bf g} $1.5 \,{\rm \upmu m}$, and {\bf j} $1 \,{\rm \upmu m}$. The yellow and cyan sections and white arrows symbolize the rotating ring, the immobile pillar, and the orientation of the metasurface, respectively. 
    {\bf b}, {\bf e}, {\bf h}, {\bf k} Corresponding SEM images. Scale bars: $5 \,{\rm \upmu m}$. 
    {\bf c}, {\bf f}, {\bf i}, {\bf l} Optical microscopy images of the metarotors rotating under linearly polarized light captured at times $0\,{\rm ms}$, $500\,{\rm ms}$, $1000\,{\rm ms}$, $1500\,{\rm ms}$, and $2000\,{\rm ms}$. Scale bars: $10 \,{\rm \upmu m}$.
    }
    \label{FigS6}
\end{suppfigure}

\begin{suppfigure}[p]
    \centering
    \includegraphics[width=\textwidth]{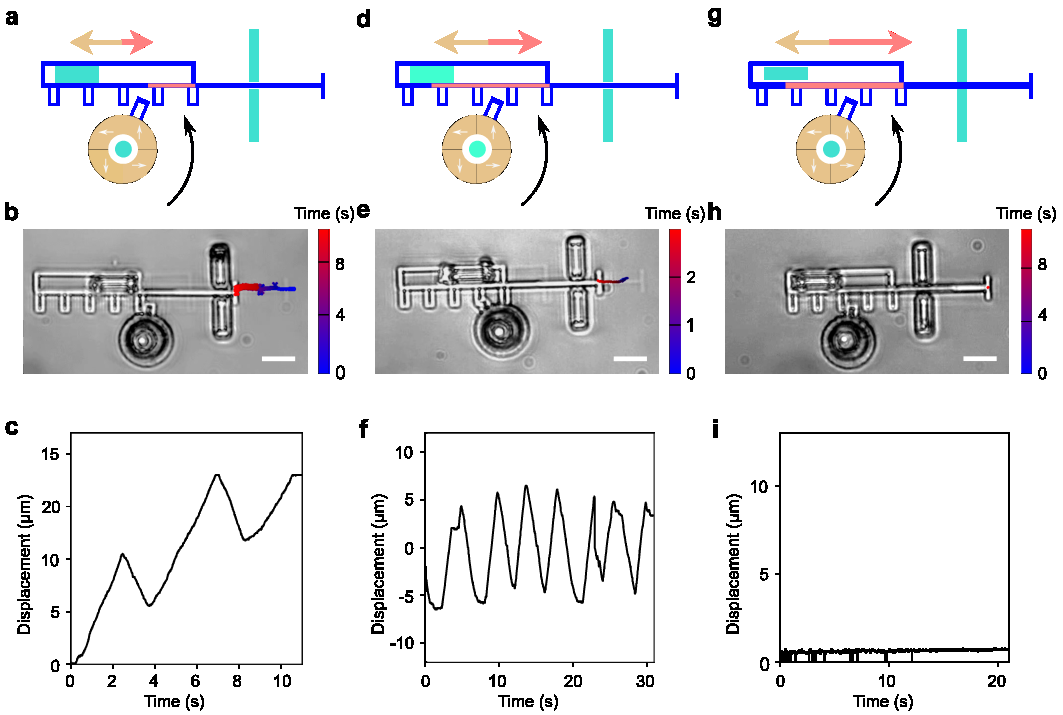}
    \caption{\textbf{Balance between the strength of the metasurfaces in the racks and in the metarotors.} {\bf a,d,g} Schematic illustration of rack and pinion metamachines with varying numbers of meta-atoms embedded in the rack: {\bf a} 5, {\bf d} 12, and {\bf g} 22 meta-atoms. The metasurface segments are illustrated in red and yellow, and the forces they induce onto the motor under linearly polarized light are depicted by white arrows. Movable gear and rack parts are shown in blue, and immobile parts in turquoise. The forces from the motor and rack on the rack are indicated by yellow and red arrows, respectively. {\bf b,e,h} Optical microscopy images of the rack and pinion design under linearly polarized light. The movement of the rack is tracked. {\bf c,f,i} Rack displacement vs time. {\bf a-c} When the force applied by the motor to the rack is greater than the force exerted by the rack itself, the motor drives the rack continuously to the left until it halts. {\bf e-f} When the force applied by the motor to the rack is approximately equal to the force exerted by the rack itself, an oscillatory back-and-forth motion occurs. {\bf g-i} Conversely, when the force exerted by the rack itself is significantly greater than the force applied by the motor, the motion of the rack itself will be inhibited. Scale bars: $10\,{\rm \upmu m}$. 
        }
    \label{FigS7}
\end{suppfigure}

\begin{suppfigure}[p]
    \centering
    \includegraphics[width=\textwidth]{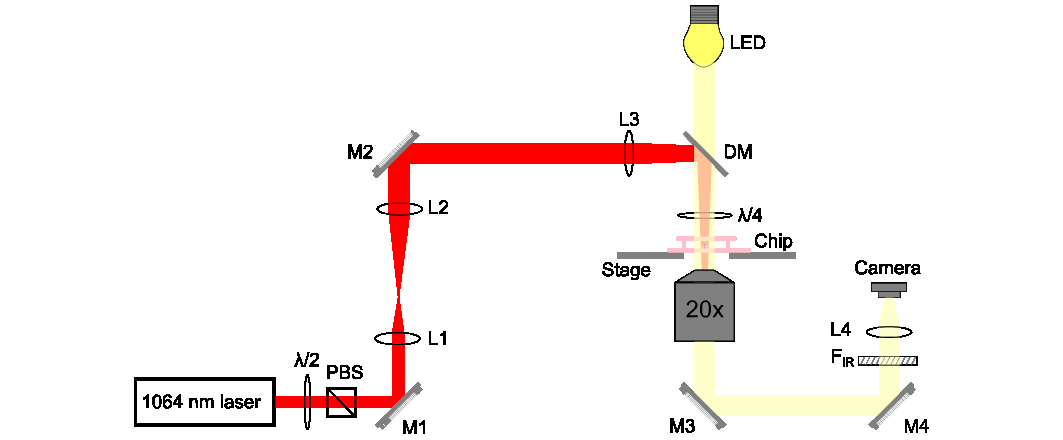}
    \caption{\textbf{Optical setup}. The beam from a $1064\,{\rm nm}$ continuous-wave laser is expanded to approximately $1\,{\rm cm}$ in diameter using a telescope system (L1, L2) and then focused through a lens (L3) to achieve a spot size of about $300\,{\rm \upmu m}$. The output power of the laser to the chip is regulated manually using the combination of a half-wave plate ($\lambda/2$) and a polarizing beam splitter (PBS) after the laser. By employing a combination of half-wave ($\lambda/2$) and quarter-wave ($\lambda/4$) plates after a dichroic mirror (DM), the polarization state can be adjusted to linear polarization or circular polarization with either right or left-handedness. The chip containing micromotors or micromachines within a thin liquid sample cell is placed on the stage, and the laser is directed onto it from above. The movements of the micromotors or micromachines are captured through imaging using a $20\times$ objective and a CMOS camera. The infrared filter is positioned in front of the camera to prevent the $1064\,{\rm nm}$ light from reaching the camera. PBS: Polarizing beamsplitter; M: Mirror; L: lens; DM: Dichroic mirror; F$_I$$_R$: Infrared filter; $\lambda/2$: Half-wave plate; $\lambda/4$: Quarter-wave plate.
    }
    \label{FigS8}
\end{suppfigure}

\clearpage

\textcolor{blue}{Supplementary Video 1}: Animation (left panel) and brightfield video (right panel) of the movement of a $16\,{\rm \upmu m}$ diameter micromotor under the illumination of an $88.5\,{\rm \upmu W \, \upmu m^{-2}}$ linearly polarized $1064\,{\rm nm}$ laser. The video is played at $0.15\times$ speed.

\textcolor{blue}{Supplementary Video 2}: Scanning electron microscopy images (top panels) of four micromotors with $16\,{\rm \upmu m}$ diameter, each embedded with different amounts of meta-atoms: $22$, $29$, $36$, $55$ in a quarter area, and corresponding brightfield videos (bottom panels) of their movement under the illumination of an $88.5\,{\rm \upmu W \, \upmu m^{-2}}$ linearly polarized $1064\,{\rm nm}$ laser. The video is played in real time.

\textcolor{blue}{Supplementary Video 3}: Scanning electron microscopy image of a $16\,{\rm \upmu m}$ diameter micromotor (left panel) and brightfield videos (other panels) of their movement under the illumination of a linearly polarized $1064\,{\rm nm}$ laser with different intensities:  $8.75\, {\rm \upmu W \, \upmu m^{-2}}$, $12.75\, {\rm \upmu W \, \upmu m^{-2}}$,  $30.25\, {\rm \upmu W \, \upmu m^{-2}}$,  $48.0\, {\rm \upmu W \, \upmu m^{-2}}$, $70.75\, {\rm \upmu W \, \upmu m^{-2}}$, $88.5\, ,{\rm \upmu W \, \upmu m^{-2}}$. The video is played in real time.

\textcolor{blue}{Supplementary Video 4}: Scanning electron microscopy images (tops panels) of four micromotors with $16\,{\rm \upmu m}$ diameter with different gap sizes ($2.5$, $2.0$, $1.5$, $1.0\,{\rm \upmu m}$) between the centeral pillar and ring-shaped structure, and corresponding brightfield images (bottom panels) of their movement under the illumination of an $88.5\,{\rm \upmu W \, \upmu m^{-2}}$ linearly polarized $1064\,{\rm nm}$ laser. The video is played in real time.

\textcolor{blue}{Supplementary Video 5}: Scanning electron microscopy image (left panel) of an $8\,{\rm \upmu m}$ diameter micromotor, and corresponding brightfield video (right panel) of its movement under the illumination of an $88.5\,{\rm \upmu W \, \upmu m^{-2}}$ linearly polarized $1064\,{\rm nm}$ laser. The video is played in real time.

\textcolor{blue}{Supplementary Video 6}: Scanning electron microscopy images and corresponding brightfield videos of gear trains with different numbers of gears: $1$, $2$, $3$, $4$ and $5$, powered by driving metagears under the illumination of an $88.5\,{\rm \upmu W \, \upmu m^{-2}}$ linearly polarized $1064\,{\rm nm}$ laser. The video is played in real time.

\textcolor{blue}{Supplementary Video 7}: Scanning electron microscopy images and corresponding brightfield videos of gear trains with different configurations and of microdrones with extended arms, powered by driving metagears under the illumination of an $88.5\,{\rm \upmu W \, \upmu m^{-2}}$ linearly polarized $1064\,{\rm nm}$ laser. The video is played in real time.

\textcolor{blue}{Supplementary Video 8}: Scanning electron microscopy image (left panel) of four micromotors with $16\,{\rm \upmu m}$ diameter, each embedded with meta-atoms oriented in different configurations, and corresponding brightfield video (right panel) of their movement under the illumination of an $88.5\,{\rm \upmu W \, \upmu m^{-2}}$ $1064\,{\rm nm}$ laser with dynamically changing polarization, the colored lines represents the tracked trajectory over the last $3\,{\rm s}$. The video is played in real time.

\textcolor{blue}{Supplementary Video 9}: Scanning electron microscopy image (left panel) and corresponding brightfield video of gear trains with different diameter of passive gears: $10\,{\rm \upmu m}$, $16\,{\rm \upmu m}$, powered by powered by optical metamaterials under the illumination of an $88.5\,{\rm \upmu W \, \upmu m^{-2}}$, circularly polarized $1064\,{\rm nm}$ laser. The video is played in real time.

\textcolor{blue}{Supplementary Video 10}: Scanning electron microscopy image (top panel) and corresponding brightfield video (bottom panel) of a microscopic rack and pinion machine operated by motor metagears under the illumination of an $88.5\,{\rm \upmu W \, \upmu m^{-2}}$, circularly polarized $1064\,{\rm nm}$ laser. The movement direction of the machine can be changed by light polarization. The video is played in real time.

\textcolor{blue}{Supplementary Video 11}: Scanning electron microscopy images (top panels) and corresponding brightfield videos (bottom panels) of microscopic machines that can only move linearly left and right under the illumination of an $88.5\,{\rm \upmu W \, \upmu m^{-2}}$ linearly polarized $1064\,{\rm nm}$ laser. The left movement is a rack and pinion machine operated by motor metagears, and the right movement is a rack powered by the meta-atoms on the rack. The video is played in real time.

\textcolor{blue}{Supplementary Video 12}: Scanning electron microscopy images (top panels) and corresponding brightfield videos (bottom panels) of the movement of microscopic rack and pinion micromachines under the illumination of an $88.5\,{\rm \upmu W \, \upmu m^{-2}}$ linearly polarized $1064\,{\rm nm}$ laser. The movement of the micromachines is based on the balance between the applied force from the meta-atoms on the metagear ($F_{\text{gear}}$) and the rack ($F_{\text{rack}}$). The micromachine in the left panel moves left until it is blocked, as $F_{\text{gear}} > F_{\text{rack}}$. The micromachine in the middle panel performs an oscillating motion in both directions, with $F_{\text{gear}} \approx F_{\text{rack}}$. The micromachine in the right panel moves right and eventually gets blocked, as $F_{\text{gear}} < F_{\text{rack}}$. The video is played in real time.

\textcolor{blue}{Supplementary Video 13}: Scanning electron microscopy image (top panel) and corresponding brightfield video (bottom panel) of a microscopic rack and pinion machine with oscillating motion to the left and right under the illumination of an $88.5\,{\rm \upmu W \, \upmu m^{-2}}$ linearly polarized $1064\,{\rm nm}$ laser. Two gold mirrors are connected to the rack and move together with it, functioning to reflect light positionally. The video is played in real time.

\end{document}